\documentclass[aps,prl,twocolumn,superscriptaddress,showpacs]{revtex4-1}
\usepackage{amsmath,amssymb}
\usepackage{graphicx}
\usepackage{natbib}
\usepackage{color}

\begin{document}

\title{Coherently Enhanced Wireless Power Transfer}

\author{Alex~Krasnok}
\affiliation{Department of Electrical and Computer Engineering, The University of Texas at Austin, Austin, Texas 78712, USA}

\author{Denis~G.~Baranov}
\affiliation{Department of Physics, Chalmers University of Technology, 412 96 Gothenburg, Sweden}
\affiliation{Moscow Institute of Physics and Technology, Dolgoprudny 141700, Russia}

\author{Andrey~Generalov}
\affiliation{Department of Electronics and Nanoengineering, Aalto University, 02150 Espoo, Finland}

\author{Sergey~Li}
\affiliation{ITMO University, St. Petersburg 197101, Russia}

\author{Andrea~Al\`u}
\email[]{alu@mail.utexas.edu}
\affiliation{Department of Electrical and Computer Engineering, The University of Texas at Austin, Austin, Texas 78712, USA}

\begin{abstract}
Extraction of electromagnetic energy by an antenna from impinging external radiation is at the basis of wireless communications and power transfer (WPT). The maximum of transferred energy is ensured when the antenna is conjugately matched, i.e., when it is resonant and it has an equal coupling with free space and its load, which is not easily implemented in near-field WPT. Here, we introduce the concept of coherently enhanced wireless power transfer. We show that a principle similar to the one underlying the operation of coherent perfect absorbers can be employed to improve the overall performance of WPT and potentially achieve its dynamic control. The concept relies on coherent excitation of the waveguide connected to the antenna load with a backward propagating signal of specific amplitude and phase. This signal creates a suitable interference pattern at the load resulting in a modification of the local wave impedance, which in turn enables conjugate matching and a largely increased amount of energy extracted to the waveguide. We develop an illustrative theoretical model describing this concept, demonstrate it with full-wave numerical simulations for the canonical example of a dipole antenna, and verify it experimentally in both near-field and far-field regimes.
\end{abstract}

\maketitle

The antenna is a key element in wireless technology, including communications and power transfer~\cite{Balanis}. The first antennas emerged at the time of the discovery of electromagnetic waves by H.~Hertz in 1888 and, since then, this technology has been progressing continuously. A plethora of optimized antennas have been invented for radio, microwave, THz and optical frequencies, where they have become irreplaceable elements for quantum optics and interconnections on a chip~\cite{AluBook, LukasNovotny2012}. 

While wireless communications are well established, wireless power transfer (WPT), proposed in the beginning of 20th century by N.~Tesla~\cite{Brown1984}, has been experiencing a rebirth in recent years, caused by demonstrations that the WPT efficiency, i.e., the ratio of energy received by an antenna over the total amount of emitted energy, can be drastically enhanced in the so-called \textit{near-field WPT} regime, when the power is transferred via resonant coupling~\cite{Kurs2007}. Transfer over a distance of 2~m with 45$\%$ efficiency in the kHz range via strongly coupled magnetic resonances between two metallic coils has been recently achieved~\cite{Kurs2007}, giving rise to research on ways of using this effect for several technologies, in which recharging batteries without cables and wires would be of great importance. Examples include electric vehicles, implanted medical devices~\cite{ImplantPNAS}, and consumer electronics~\cite{Kim2013, Mei2014, Kim2012}. Significant research efforts have been recently devoted in exploring ways to achieve high WPT efficiency~\cite{Song2017}, optimizing the resonators' geometry~\cite{Song2016}, the surrounding materials~\cite{Wang2011, Urzhumov, Kim2013}, and their relative arrangement~\cite{Kurs2007}. In addition, great progress has been recently made to ensure the overall robustness of near-field WPT systems as a function of variations in the environment and background, using active systems, nonlinearity and feedback~\cite{Assawaworrarit2017, Radi2017}. Since near-field WPT systems require that receiving and transmitting antennas are resonant and, thus, restrict their minimal sizes and relative distances, they are quite impractical for modern WPT systems, which mostly rely on far-field non-resonant WPT technology~\cite{Song2017}.
 
\begin{figure}[b]
	\includegraphics[width=.99\columnwidth]{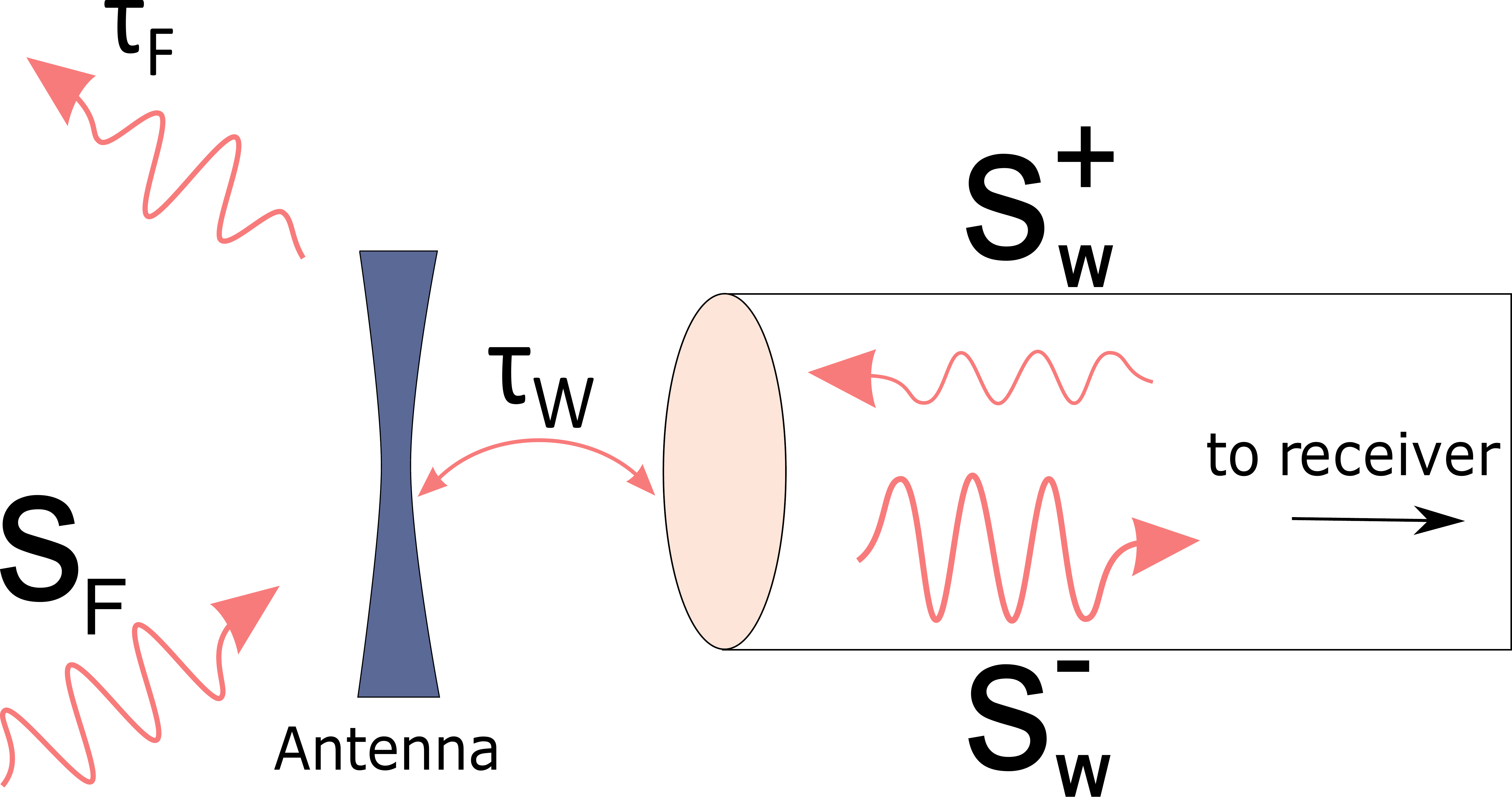}
	\caption{Sketch of coherently enhanced wireless energy transfer. The incident radiation ($s_F$) excites the antenna, which re-radiates back into free space at a rate $1/\tau_F$ and couples to waveguide at a rate $1/\tau_{\rm w}$ creating a forward-propagating waveguide mode ($s_{\rm w}^-$). Coherent excitation of the outcoupling waveguide with a backward propagating guided mode ($s_{\rm w}^+$) with specific amplitude and phase can retune the matching condition and largely improve the WPT performance. Figure-of-merit of this model system, suitable for both far- and near-field WPT systems, is the energy balance $|s_{\rm w}^-|^2-|s_{\rm w}^+|^2$.}
	\label{Fig1}
\end{figure}

\begin{figure*}
	\includegraphics[width=0.99\textwidth]{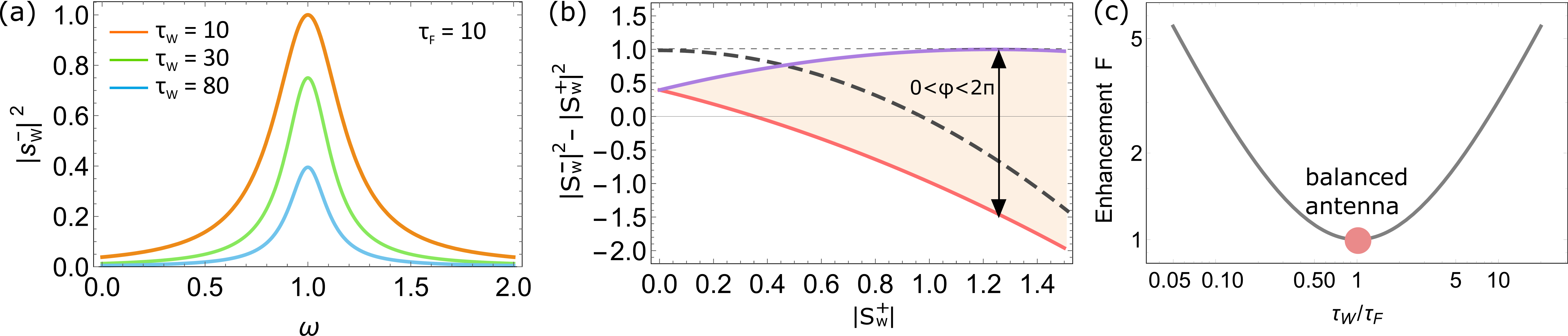}
	\caption{Analytical modeling of coherently enhanced wireless power transfer. (a)~Spectrum of the transferred energy ($|s_{\rm w}^-|^2$) without the back-propagating signal for antennas with different waveguide coupling. (b)~Energy balance for a resonant matched (gray dashed curve) and resonant mismatched antenna (color solid curves). Filled area indicates the available values for the energy benefit. (c)~Enhancement factor $F$ defined as the ratio of maximum achievable transferred energy assisted by the back-propagating signal to the transferred energy $|s_{\rm w}^-|^2$ without coherent illumination as a function of $\tau_{\rm w} / \tau_F$.}
	\label{Fig2}
\end{figure*}

In this letter, we propose a conceptually different approach to realize robust WPT systems, relying on the local control of the wave impedance offered by interference phenomena. Recent research has shown that electromagnetic processes such as absorption and scattering may be effectively controlled via coherent spatial and temporal shaping of the incident electromagnetic field. For example, a coherent perfect absorber (CPA) is a linear electromagnetic system in which perfect absorption of radiation is achieved with two or more incident coherent waves, creating constructive interference inside an absorbing structure~\cite{Chong2010, Chong2011, Zhang2012, BaranovAlexKrasnok2017}. Similar principles also allow developing linear logic gates~\cite{Fang2015} and pattern recognition setups~\cite{Zh-2017}, in which interference between two input waves is the enabling factor. In this Letter, we show that this approach can be employed to improve and control dynamically the matching of receiving antennas, and as a result enhance the overall performance of WPT systems and retune the system due to the changes in the channel link between transmitter and receiver. More specifically, we demonstrate that there is a possibility to improve the receiving efficiency of an antenna by \textit{coherent excitation} of the outcoupling waveguide with a backward propagating signal of specific amplitude and phase, Fig.~\ref{Fig1}. This signal creates a specific interference pattern in the system that results in optimal wave impedance at the feed location, maximizing the energy transferred to the receiving antenna from free space. We develop an illustrative analytical model predicting this effect, demonstrate it in full-wave numerical simulations, and in a microwave experiment.

In order to demonstrate the effect of coherently enhanced power transfer, we develop a theoretical model on the basis of temporal coupled mode theory~\cite{Haus, Fan2003}, which applies for both near-field and far-field WPT systems. The system of a waveguide-coupled antenna is schematically shown in Fig.~\ref{Fig1}. The antenna is excited by a field $s_F$ created by a transceiver, which can be either free-propagating or evanescent. The antenna couples to a waveguide with amplitude $s_{\rm w}^-$, which carries the out-coupled energy to the receiver. To model the structure, we will assume that the antenna has a single resonance at the operating frequency $\omega_0$. The dynamics of the resonance mode amplitude $a$ can be described by the equation~\cite{Haus, Fan2003}
\begin{equation}
\frac{{da}}{{dt}} = (i{\omega _0} - 1/\tau )a + \left\langle {{\kappa ^*}|{s_ + }} \right\rangle,
\end{equation}
where $\tau$ is the mode damping time,  $\left| \kappa \right\rangle =(\kappa_{\rm w}, \kappa_F)$ is the coupling constants vector, and $\left| s_+ \right\rangle =(s_{\rm w}^+, s_F)$ is the input amplitudes vector. The output amplitude vector $\left| {{s_ - }} \right\rangle $, on the other hand, is related to the input vector and the resonance amplitude via
\begin{equation}
\left| {{s_ - }} \right\rangle = \hat C \left| {{s_ + }} \right\rangle + a\left| \kappa \right\rangle,
\end{equation}
where $\widehat C$ is the \emph{direct scattering matrix}, which reflects the direct pathway between input and output channels.

The two crucial parameters that determine the system response are $\tau_{\rm w}$ and $\tau_F$, being the antenna decay times into the waveguide mode and into radiation in free space, respectively. If the antenna does not exhibit large dissipative losses (which is a reasonable assumption in the microwave region), the total decay time is given by $1/\tau = 1/{\tau_{\rm w}} + 1/{\tau _F}$. The coupling constants are related to the corresponding decay times as ${\kappa _{\rm w}} = \sqrt {2/{\tau_{\rm w}}} $ and ${\kappa _F} = \sqrt {2/{\tau _F}} $~\cite{Fan2003}. The direct scattering matrix, in turn, satisfies $\widehat C{\left| \kappa \right\rangle ^*} = - \left| \kappa \right\rangle $~\cite{Fan2003}.

Eqs.~(1) and (2) establish the relation between the continuum of free-space modes and the discrete waveguide mode. The goal of efficient WPT is to enhance the amplitude of the back scattered waveguide mode, denoted by $s_{\rm w}^-$ in Fig.~\ref{Fig1}. Therefore, for the sake of simplicity, we reduce the direct scattering matrix to a scalar $c$, assuming that the coupling of the antenna with free-space modes leading to additional radiation losses is taken into account by $\tau _F$. Such simplification does not undermine our model. Furthermore, for a microwave waveguide or coaxial cable it may be safely assumed that there is no direct scattering from the waveguide into free space (i.e., this interaction is mostly mediated by the antenna). Therefore, $c = - {\kappa_{\rm w}}/\kappa_{\rm w}^*=-1$, and the resulting amplitude of the reflected mode that carries the transferred energy is given by
\begin{equation}
{s_{\rm w}^- } = c{s_{\rm w}^+ } + {\kappa_{\rm w}}\frac{{{\kappa _{\rm w}}{s_{\rm w}^+ } + {\kappa _F}{s_F}}}{{i(\omega - {\omega _0}) + 1/\tau }}.
\label{eq3}
\end{equation}
With $s_F=0$ this expression yields the reflection s-parameter ($s_{11}$) that can be easily measured experimentally or calculated numerically.

Without the back-propagating waveguide mode ($s_{\rm w}^+=0$), this expression results in a conventional Lorentzian spectrum for the transferred energy $|s_{\rm w}^-|^2$, reaching a maximum at the resonant frequency $\omega_0=1$, see Fig.~\ref{Fig2}(a), where the calculations are presented for amplitude of the impinging free space mode $|s_F|=1$. This maximum can be easily found from Eq.~\ref{eq3} and it is equal to $4{\tau_{\rm w}}{\tau _F}/{({\tau_{\rm w}} + {\tau _F})^2}$. The optimal condition for coupling of the incident free space mode into the waveguide is known as the \emph{critical coupling}, or conjugate matching condition~\cite{Balanis}, and it is achieved when $\tau_{\rm w}=\tau_F$. This condition maximizes the amount of energy transmitted into the waveguide at resonance. While it is always possible to design the antenna so that the load resistance balances the radiation resistance to satisfy this condition, in practice as the environment, or generally the channel connecting transmitter and receiver, change, the load needs to be retuned to satisfy the conjugate matching condition. In the following, we show that it is possible to maximize transmission without affecting the coupling parameters, and retune the conjugate matching condition in real time by coherent illumination of the antenna from the loading port, similarly to the concept of coherent perfect absorption, when critical coupling for absorption in an unbalanced system is restored by an impinging coherent wave  with proper phase and amplitude~\cite{Chong2010, BaranovAlexKrasnok2017}.

We now inspect how an auxiliary back-propagating wave impinging from the waveguide port at the load of the antenna can modify the input-output characteristic of the system. In Fig.~\ref{Fig2}(b) we show the \textit{energy balance} in transferred energy, defined as the total received energy $|s_{\rm w}^-|^2$ minus the energy spent in the auxiliary signal $|s_{\rm w}^+|^2$ that is sent from the waveguide port to the load, as a function of the amplitude $|s_{\rm w}^+|$ for different values of the  relative phase $\varphi=\text{Arg}(s_{\rm w}^+/s_F)$. We chose the energy balance as a figure-of-merit since it allows to treat far-field and near-field WPT systems on an equal footing. A positive (negative) energy balance indicates that the antenna receives (radiates) more energy than radiates (receives). More importantly, if the energy balance exceeds the receiving energy $|s_{\rm w}^-|^2$ for $|s_{\rm w}^+|^2=0$ (no auxiliary signal), it signifies more favorable coherently assisted performance of the WPT system. For an antenna  that is already conjugate matched (gray dashed line), as expected, the maximum power transfer arises when the auxiliary signal from the port is absent, since the signal would only detune the already optimal matching condition. This curve does not depend on the relative phase since the signal $s_{\rm w}^+$ does not reflect back and, hence, does not alter the total receiving energy $|s_{\rm w}^-|^2$. However, for a mismatched antenna (shaded region in the plot, bounded between two colored lines) the input-output characteristic exhibits a peculiar behavior. The shaded region marks all achievable values of the energy balance for phase differences spanning the whole range between 0 and 2$\pi$, indicating that for a suitable combination of phase and amplitude it is possible to achieve perfectly matched regime (the energy balance is positive and larger than in the case of no coherent excitation), even after subtracting the energy carried by the back-propagating signal. In particular, for a certain phase difference the amount of extracted power (purple curve) reaches the ideal value of a resonant critically coupled antenna. This coherent illumination restores the critical waveguide-antenna coupling, without having to retune the antenna load impedance. The specific value of optimal phase that ensures critical coupling depends on the direct scattering matrix $\widehat C$; if we allow direct pathways between the free space and the waveguide, or a reactive mismatch of the load, the phases of reflected and transmitted waves would be different and the optimal phase difference would change.

\begin{figure}
	\includegraphics[width=.45\textwidth]{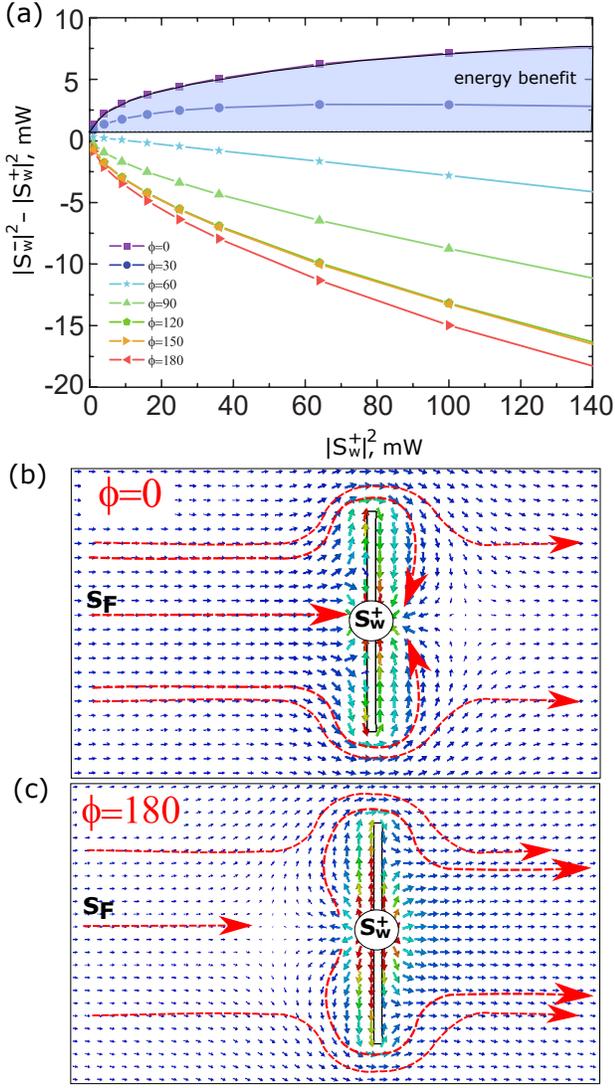}
	\caption{Numerical simulations of the proposed approach of coherently enhanced WPT for the canonical example of a dipole antenna. (a)~Net extracted energy ($|s_{\rm w}^-|^2-|s_{\rm w}^+|^2$) as a function of the auxiliary signal intensity ($|s_{\rm w}^+|^2$) for different relative phase values. The black dashed line shows the energy level received by the dipole antenna without auxiliary signal. Filled area indicates the region of energy benefit, where the coherently assisted energy balance exceeds that for $s_{\rm w}^-=0$.  (b), (c)~Poynting vector distribution around the microwave antenna with the auxiliary signal of amplitude $|s_{\rm w}^+|=12$ and relative phases of 0~deg (b) and 180~deg (c).}
	\label{Fig3}
\end{figure}

To better illustrate the capabilities of this approach, we show in Fig.~\ref{Fig2}(c) the enhancement factor $F$ defined as the ratio of maximum achievable transferred energy assisted by the auxiliary mode to the transferred energy $|s_{\rm w}^-|^2$ without coherent illumination as a function of $\tau_{\rm w} / \tau_F$. It is seen that in the critically coupled case the enhancement factor equals 1, since the amount of transferred energy cannot be increased by coherent waveguide excitation. For mismatched antenna, on the other hand, the coherence-assisted enhancement factor increases with deviation of $\tau _{\rm w}/\tau _F$ from 1, and therefore can have large values for strongly mismatched antennas. In the radio frequency range, mismatched and non-resonant antennas are common in the context of various wireless charging devices, in which the antenna, of cm scale length, is far off-resonant, since the radiation of the wireless transfer platform lying in the 100~kHz range~\cite{Song2017}.  Our approach allows tuning the antenna to the optimal condition through wave interference. In addition, in many situations the matching condition can be easily detuned by changes in the background environment, for instance by neighboring parasitic reflections, or simply by a changing distance of the transmitter. Our approach offers a viable solution to real-time retuning of the antenna by sending a signal from the receiving port to modify the local wave impedance at the load.

\begin{figure*}
\includegraphics[width=.8\textwidth]{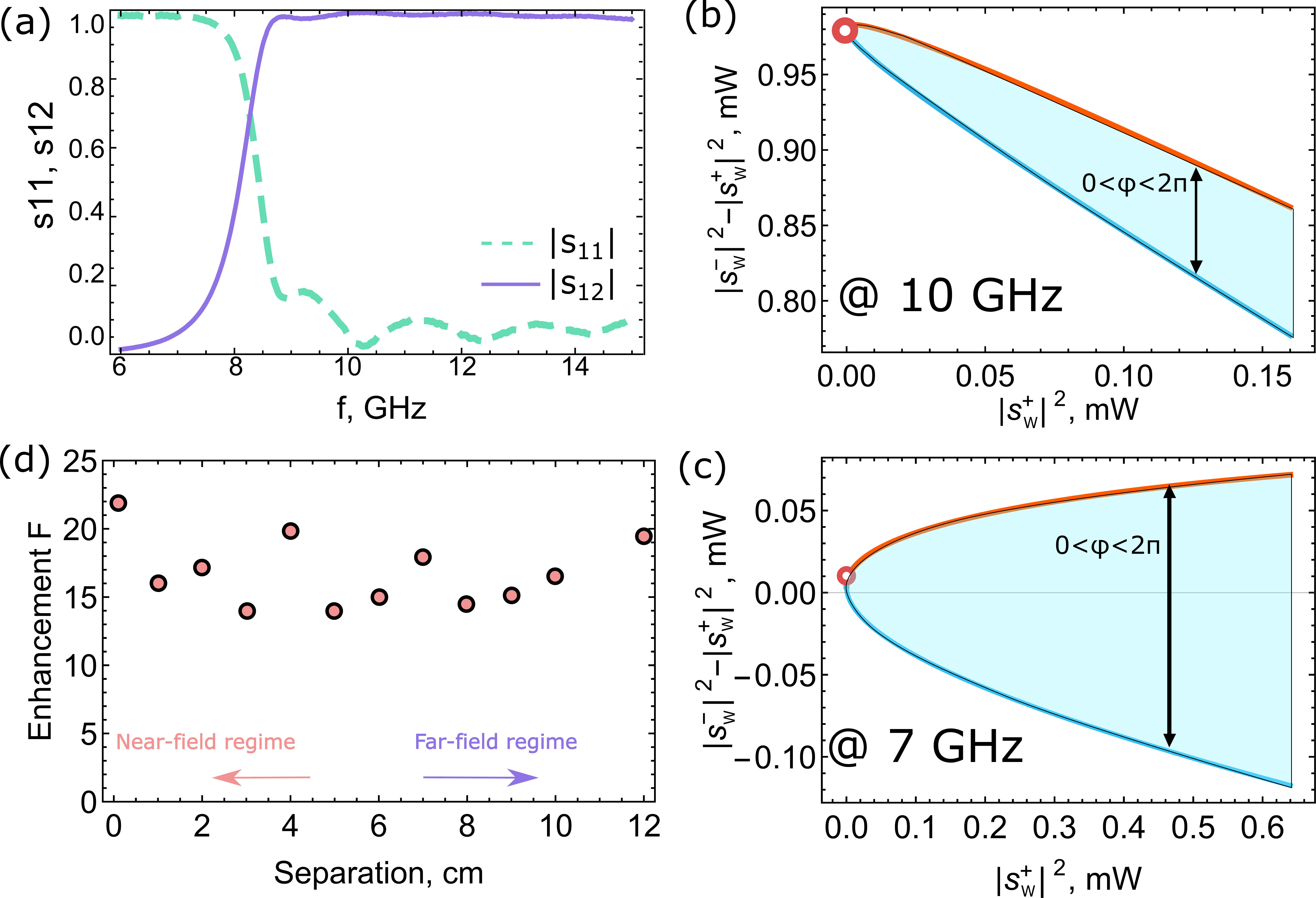}
	\caption{Experimental verification of coherently enhanced wireless energy transfer. (a)~Measured spectrum of $ s_{11}$ and $ s_{22}$ absolute values for transmitting and receiving antennas placed next to each other. Inset: Photo of the experimental setup. Two coaxial cables are attached to the vector network analyzer ports and terminated with waveguide to coaxial adapters at the free ends. One port serves as the transmitter, while the other serves as the power receiver. (b,c)~The net extracted energy $| s_{\rm w}^-|^2 - | s_{\rm w}^+|^2$ vs power of the additional signal $| s_{\rm w}^+|^2$ for two different frequencies. The shaded area corresponds to all values of the relative phase between the external (default or main) signal and the auxiliary mode. The power of 1~mW is fed to port 1 in both cases. (d)~Dependence of the energy transfer enhancement factor $F$ on the distance between the WCAs at fixed frequency of 7~GHz.}
	\label{Fig4}
\end{figure*}

To verify the predictions of the analytical model, we carried out FDTD simulations of coherently enhanced WPT using the CST Microwace Studio. In our model, a coaxial cable was coupled to a non-resonant dipole antenna with 5~cm arm-length. Microwave radiation at 1.36~GHz (with a wavelength of 22~cm) from the open end of a rectangular waveguide separated by a distance of 40~cm from the antenna was used as the incident radiation ($s_F$). A coherent auxiliary signal ($s_{\rm w}^+$) was launched into the coaxial cable in order to enable the interference. The resulting dependence of the energy balance shown in Fig.~\ref{Fig3}(a) reproduces the theoretical predictions. We note that the relative phase denoted here as $\phi$ is calculated for the ${\rm Arg}(s_{\rm w}^+)$ and ${\rm Arg}(s_F)$ taken at different points (at the dipole antenna center and at the rectangular waveguide end, respectively) and, hence, depends on their relative distance.

To provide more insight into the observed behavior, we present in Figs.~\ref{Fig3}(b,c) the Poynting vector distribution around the antenna considering the coherent excitation with relative phases of 0~deg [Fig.~\ref{Fig3}(b)] and 180~deg [Fig.~\ref{Fig3}(c)]. It is evident that, when the antenna is coherently excited by a wave of appropriate amplitude and phase, it enables largely enhanced antenna aperture that collects more energy into the coaxial cable from the propagating wave, Fig.~\ref{Fig3}(b). On the other hand, when the relative phase is not well chosen, the external wave ($s_F$) enables enhanced radiation of the dipole antenna into free space.

Finally, we perform a proof-of-principle experiment to demonstrate the feasibility of coherently enhanced WPT. For this purpose, we put together a microwave two-port system shown in the inset of Fig.~\ref{Fig4}(a). Each coaxial cable is connected to a vector network analyzer and terminated with a waveguide to coaxial adapter (WCA) that transforms electrical currents to propagating electromagnetic radiation. Fig.~\ref{Fig4}(a) presents the absolute values of measured $s_{11}$ and $s_{12}$ scattering parameters for arrangement of the system with the WCA faces next to each other. The WCAs have cutoff frequency around 8~GHz, which manifests itself in high reflection ($s_{11}$) at frequencies below 8~GHz. Thus, the WCA operates as a mismatched antenna for frequencies below 8~GHz and an almost matched antenna for frequencies above 8~GHz. The port 1 creates free space radiation s$_F$, a part of which (s$^-_{\rm w}$) then receives by port 2. The port 2 is excited by the auxiliary signal (s$^+_{\rm w}$) of variable amplitude and phase.

The resulting $s$-parameter spectra allow to observe coherently enhanced energy transfer. The dependence of the net extracted energy $|s_{\rm w}^-|^2 - |s_{\rm w}^+|^2$ versus power of the additional excitation $|s_{\rm w}^+|^2$ for the whole set of relative phases $0<\phi<2\pi$ presented in Fig.~\ref{Fig4}(b) and (c) highlights the different behavior of the system in various spectral regions. At 10~GHz, where reflection by each WCA back to the coaxial cable is very low, excitation with the auxiliary signal barely improves the energy transfer, Fig.~\ref{Fig4}(b). At 7~GHz, however, the situation is strikingly different: high reflection by the WCA enables order of magnitude enhancement of the transferred power, Fig.~\ref{Fig4}(c). This enhancement becomes possible due to a small transmission from port 1 to port 2 even at 7~GHz, which can be increased by interference with the auxiliary signal.

To further inspect the opportunities offered by coherently enhanced energy transfer, we study the input-output behavior of our setup for different distances between the WCAs. Fig.~\ref{Fig4}(d) shows the enhancement factor $F$ versus the distance at the fixed frequency of 7~GHz (wavelength of 4.28~cm), where the WCAs operate as largely mismatched antennas. At each distance, the maximum of $F$ is achieved for a specific relative phase between the two signals, which is determined by complex values of $s$-parameters for the given arrangement of the setup. An enhancement of the received energy around 20 can be achieved at this frequency for the whole range of distances (0--12~cm, which covers both near- and far-field regimes), highlighting the great versatility of the present approach for both near and far field WPT systems, and the possibility or realizing robust WPT independent of variations of the distance between transmitter and receiver.

We envision that in a practical WPT device, one may control in real time the amplitude and phase of the auxiliary signal to retune the antenna as a function of changes in the environment, temperature changes in the load, and distance of the transmitter. We propose the employment of an adaptive filter in the receiver circuit that monitors the transferred power and self-adjusts the amplitude and phase of the auxiliary signal such that it maximizes in real-time the transferred power.

In conclusion, we have shown that coherent signals sent from the receiving port of a WPT system can largely enhance and control the power transfer efficiency. This additional signal creates a tailored interference in the system, modifying the local wave impedance at the antenna load, thus enabling conjugate matching and critical coupling even if the antenna itself is largely mismatched, resulting in increased amount of energy extracted to the waveguide from free space. We have developed an illustrative analytical model predicting this effect and demonstrated it in numerical simulations and in a microwave experiment. Our approach of coherently enhanced WPT can be applied for the development of efficient wireless power transfer systems with robust operation in rapidly changing environments, as common in practical situations and setups.

The authors are grateful to Prof.~Constantin~Simovski and Prof.~Sergei~Tretyakov for their expert advice and helpful criticism during the elaboration of this work. This work was partially supported by the Air Force Office of Scientific Research  and the Samsung GRO program. D.G.B. acknowledges support from the Knut and Alice Wallenberg Foundation.

%\bibliography{extractionNEW}

%

\end{document}